\documentclass[twocolumn,english,superscriptaddress,preprintnumbers,amsmath,amssymb,prb,floatfix,footinbib]{revtex4}
\usepackage[utf8x]{inputenc}
\usepackage[scaled=2]{helvet}
\usepackage[T1]{fontenc}
\usepackage[utf8]{luainputenc}
\usepackage{multirow}
\usepackage{textcomp}
\usepackage{amsmath}
\usepackage{amssymb}
\usepackage{graphicx}
\usepackage{subscript}
\usepackage{epstopdf}
\usepackage{color}
\makeatletter

\@ifundefined{textcolor}{}
{%
 \definecolor{BLACK}{gray}{0}
 \definecolor{WHITE}{gray}{1}
 \definecolor{RED}{rgb}{1,0,0}
 \definecolor{GREEN}{rgb}{0,1,0}
 \definecolor{BLUE}{rgb}{0,0,1}
 \definecolor{CYAN}{cmyk}{1,0,0,0}
 \definecolor{MAGENTA}{cmyk}{0,1,0,0}
 \definecolor{YELLOW}{cmyk}{0,0,1,0}
}

\usepackage[scaled=2]{helvet}\usepackage{amstext}

\makeatletter%%%%%%%%%%%%%%%%%%%%%%%%%%%%%% Textclass specific LaTeX commands.
\@ifundefined{textcolor}{}{\definecolor{BLACK}{gray}{0}\definecolor{WHITE}{gray}{1}\definecolor{RED}{rgb}{1,0,0}\definecolor{GREEN}{rgb}{0,1,0}\definecolor{BLUE}{rgb}{0,0,1}\definecolor{CYAN}{cmyk}{1,0,0,0}\definecolor{MAGENTA}{cmyk}{0,1,0,0}\definecolor{YELLOW}{cmyk}{0,0,1,0}}

\usepackage{dcolumn}% Align table columns on decimal point
\usepackage{bm}% bold math
\usepackage{times}% font that prx use
\usepackage{hyperref}
\hypersetup{colorlinks=true,linkcolor=red,citecolor=blue}
\makeatother

\usepackage{babel}

\makeatother

\usepackage{babel}
\begin{document}
\title{Bulk domain Meissner state in the ferromagnetic superconductor EuFe$_{2}$(As$_{0.8}$P$_{0.2}$)$_{2}$: Consequence of compromise between ferromagnetism and superconductivity}
\author{Wentao Jin}
\email{wtjin@buaa.edu.cn}
\affiliation{School of Physics, Beihang University, Beijing 100191, China}

\author{Sebastian M\"uhlbauer}
\email{Sebastian.Muehlbauer@frm2.tum.de}
\affiliation{Heinz Maier-Leibnitz Zentrum (MLZ), Technische Universit\"at M\"unchen, D-85748 Garching, Germany}

\author{Philipp Bender}
\affiliation{Physics and Materials Science Research Unit, University of Luxembourg, 162A Avenue de la Faiencerie, L-1511 Luxembourg, Grand Duchy of Luxembourg}

\author{Yi Liu}
\affiliation{College of Science, Zhejiang University of Technology, Hangzhou 310023, China}

\author{Sultan Demirdis}
\affiliation{J\"ulich Centre for Neutron Science JCNS at Heinz Maier-Leibnitz Zentrum (MLZ), Forschungszentrum J\"ulich GmbH, Lichtenbergstrasse 1, D-85747 Garching, Germany}

\author{Zhendong Fu}
\affiliation{Neutron Platform, Songshan Lake Materials Laboratory, Dongguan 523808, China}

\author{Yinguo Xiao}
\affiliation{School of Advanced Materials, Peking University Shenzhen Graduate School, Shenzhen 518055, China}

\author{Shibabrata Nandi}
\affiliation{J\"ulich Centre for Neutron Science JCNS and Peter Gr\"unberg Institut PGI, JARA-FIT, Forschungszentrum J\"ulich GmbH, D-52425 J\"ulich, Germany}

\author{Guang-Han Cao}
\affiliation{Department of Physics, Zhejiang University, Hangzhou 310027, China}

\author{Yixi Su}
\email{y.su@fz-juelich.de}
\affiliation{J\"ulich Centre for Neutron Science JCNS at Heinz Maier-Leibnitz Zentrum (MLZ), Forschungszentrum J\"ulich GmbH, Lichtenbergstrasse 1, D-85747 Garching, Germany}

\author{Thomas Br\"uckel}
\affiliation{J\"ulich Centre for Neutron Science JCNS and Peter Gr\"unberg Institut PGI, JARA-FIT, Forschungszentrum J\"ulich GmbH, D-52425 J\"ulich, Germany}
\affiliation{J\"ulich Centre for Neutron Science JCNS at Heinz Maier-Leibnitz Zentrum (MLZ), Forschungszentrum J\"ulich GmbH, Lichtenbergstrasse 1, D-85747 Garching, Germany}

\begin{abstract}
Small-angle neutron scattering (SANS) measurements are performed on the ferromagnetic superconductor EuFe$_{2}$(As$_{0.8}$P$_{0.2}$)$_{2}$ ($T\rm_{sc}=22.5\,$K) to probe the delicate interplay between ferromagnetism and superconductivity. A clear signature of large ferromagnetic domains is found below the ferromagnetic ordering temperature $T\rm_{C}$ = 18.5 K. In a small temperature interval of $\sim$ 1.5 K below $T\rm_{C}$, additional SANS signal is observed, of which the indirect Fourier transform reveals characteristic length scales in between $\sim$ 80 nm to $\sim$ 160 nm. These nanometer-scaled domain structures are identified to result from an intermediate inhomogeneous Meissner effect denoted domain Meissner state, which was recently observed on the surface of EuFe$_{2}$(As$_{0.79}$P$_{0.21}$)$_{2}$ crystals by means of magnetic force microscopy [V. S. Stolyarov $\mathit{et}$ $\mathit{al.}$, {\color{blue}{Sci. Adv. \textbf{4}, 1061 (2018)}}], ascribing to the competition between ferromagnetism and superconductivity. Our measurements clearly render the domain Meissner state as a bulk phenomenon and provide a key solution to the mystery regarding the intriguing coexistence of strong ferromagnetism and bulk superconductivity in these compounds. 

\end{abstract}
\maketitle
\renewcommand{\thefootnote}{\fnsymbol{footnote}}

The antagonistic nature of ferromagnetism (FM) and superconductivity (SC) make the coexistence of these two states of matter quite rare, as the strong exchange fields from a ferromagnet generally destroy the singlet Cooper pairing via the paramagnetic effect~\cite{Bulaevskii_85}. Theoretically, there are several scenarios in which SC and FM can reach a compromise. Firstly, by forming multidomains or a so-called \textquotedblleft cryptoferromagnetic\textquotedblright{} structure such as spiral alignment of spins, the ferromagnetic state can lower its detrimental effect on the SC~\cite{Anderson_59,Faure_05}. Secondly, the superconducting Cooper pairs can be "polarized" by the exchange fields and show non-zero momentum, leading to inhomogeneous SC in space called the Fulde-Ferrell-Larkin-Ovchinnikov (FFLO) state~\cite{Fulde_64,Larkin_65}. Thirdly, the internal magnetic field from FM may penetrate the superconductor in the form of vortices, leading to a spontaneous vortex state~\cite{Greenside_81}.

The coexistence of SC and FM was previously revealed in uranium-based heavy-fermion compounds UGe$_{2}$, URhGe, UCoGe and intercalated iron-selenide (Li,Fe)OHFeSe~\cite{Saxena_00,Aoki_00,Aoki_12,Pachmayr_15}. However, what coexists with the SC in most of these compounds is weak FM with a small ordered moment. Recently, the observation of an intriguing coexistence of bulk SC and strong FM in EuFe$_{2}$As$_{2}$-family iron pnictides (Eu122) with the superconducting critical temperature $T\rm_{sc}$ $\sim$ 22 K and ferromagnetic ordering temperature $T\rm_{C}$ $\sim$ 17 K has attracted much attention~\cite{Ren_09,Cao_11,Nandi_14,Zapf_17}. Neutron diffraction experiments have confirmed the ferromagnetic ordering of localized Eu$^{2+}$ moments with a huge moment of $\sim$ 7 $\mu$$\rm_{B}$ per Eu atom in the superconducting ground state, induced by either chemical doping into the parent compound or application of hydrostatic pressure~\cite{Jin_13,Jin_15,Jin_PhaseDiagram,Jin_Pressure,ZhouZ_19}. The unprecedentedly large saturated moment, relatively high $T\rm_{C}$ and $T\rm_{sc}$, and broad temperature range in which SC and FM coexists, make the Eu122 ferromagnetic superconductors (FMSCs) quite unique and promising for possible applications in superconducting spintronics.

Unfortunately, only very limited studies have been carried out to understand the coexistence mechanism in the Eu122 FMSCs. Through in-depth magnetometry measurements, Jiao $\mathit{et}$ $\mathit{al.}$ proposed a spontaneous vortex state as the solution for the coexistence of FM and SC in Eu(Fe$_{0.91}$Rh$_{0.09}$)$_{2}$As$_{2}$~\cite{Jiao_17}. Using magnetic force microscopy (MFM), Stolyarov $\mathit{et}$ $\mathit{al.}$ has systematically investigated how FM and SC fight with each other at nanoscale as a function of temperature in EuFe$_{2}$(As$_{0.79}$P$_{0.21}$)$_{2}$~\cite{Stolyarov_18}. Importantly, evolved from the homogeneous Meissner state (HMS) for $\mathit{T\rm_{C}}$ < $\mathit{T}$ < $\mathit{T\rm_{sc}}$, an inhomogeneous domain Meissner state (DMS) characterized by striped domains with the width of $\sim$ 100 to $\sim$ 200 nm was observed in a very narrow temperature range ($\sim$ 1 K) just below $\mathit{T\rm_{C}}$. Upon further cooling, based on the intermediate DMS, a domain vortex-antivortex state (DVS) characterized by coexisting Abrikosov vortices and ferromagnetic domains of larger size ($\mathit{l}$ $\sim$ 350 nm) is finally established well below $\mathit{T\rm_{C}}$, well consistent with expectations from the theory of magnetic domain phases in FMSCs with purely electromagnetic interaction~\cite{Devizorova_19}. However, as MFM is a surface-sensitive techniqe, confirming the existence of intermediate DMS using bulk probes is crucial for understanding the delicate interplay between SC and FM in the Eu122 FMSCs.

\begin{figure*}
\begin{center}
\includegraphics[width=0.85\textwidth]{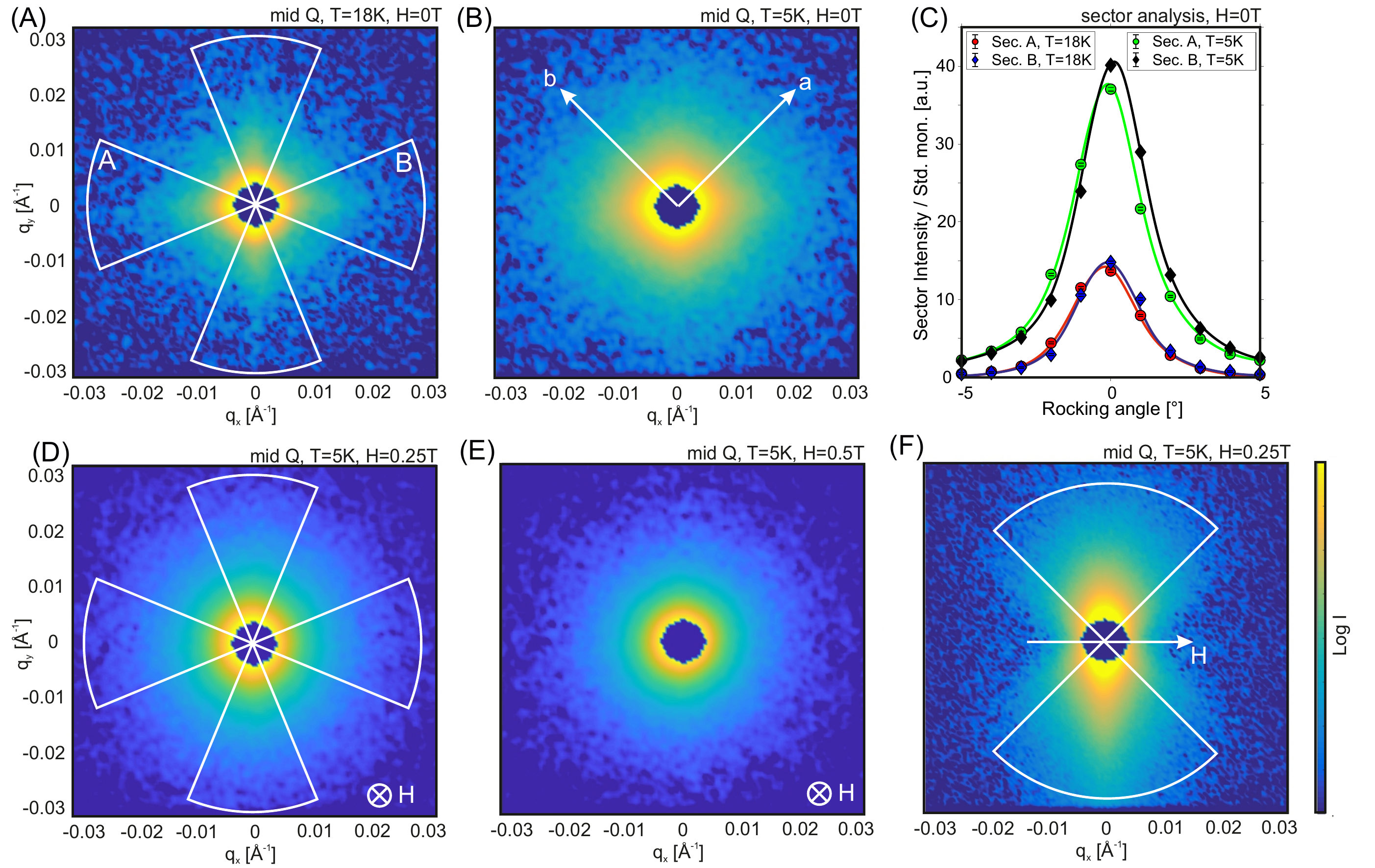}
\caption{Two-dimensional (2D) SANS patterns measured at different temperatures and magnetic fields. All 2D data shown represent the center of rocking scan of $\pm 5^{\circ}$. A high-temperature background obtained at $T=25\,$K is subtracted, and the strong signal around the direct beam is masked. (A) and (B) show the zero-field SANS patterns at $\mathit{T}$ = 18 K and $\mathit{T}$ = 5 K, respectively, with the crystallographic orientations in tetragonal notation marked by the white arrows in (B). (D) and (E) show the SANS patterns measured at $\mathit{T}$ = 5 K in a longitudinal magnetic field of $\mathit{H_{l}}$ = 0.25 T and 0.5 T parallel to the incident neutron beam along $\mathit{c}$, respectively. (F) shows the SANS pattern measured at $\mathit{T}$ = 5 K in a transverse magnetic field of $\mathit{H_{t}}$ = 0.25 T along the [1, -1, 0] direction perpendicular to the incident neutron beam. The white azimuthal sectors in (A), (D) and (F) illustrate the regions in which the intensities are integrated radially into $I(q)$ as shown in Fig. 3, and the horizontal sectors in (A) marked A and B illustrate the regions in which the intensities are integrated and shown in (C) as a function of the rocking angle.}
\label{Fig_1}
\vspace{-0.00\textwidth}
\end{center}
\end{figure*}

In this Letter, using small-angle neutron scattering (SANS) as a bulk probe, we find a clear signature of DMS in single-crystal samples of the EuFe$_{2}$(As$_{0.8}$P$_{0.2}$)$_{2}$ FMSC. In a small temperature interval of $\sim$ 1.5 K below $T\rm_{C}$, additional SANS signal revealing characteristic length scales in between $\sim$80 nm to $\sim$160 nm is observed, in addition to large and smooth ferromagnetic domains that exhibit characteristic scattering intensity of $I(q)\propto q^{-4}$ \cite{Muehlbauer_19}. These additional length scales vanish with decreasing tempature and increasing magnetic field, suggesting the important role of the intermediate DMS in the competition between SC and FM in the Eu122 FMSCs.

Single crystals of EuFe$_{2}$(As$_{0.8}$P$_{0.2}$)$_{2}$ were grown using a self-flux method~\cite{Xu_14}, and used for measurements on the diffuse scattering cold-neutron spectrometer DNS and small-angle neutron scattering instrument SANS-1 at MLZ~\cite{DNS,Muehlbauer:16}. Supporting DC magnetization measurements were performed on a Quantum Design magnetic property measurement system (MPMS) using one crystal from the same batch. Details of neutron scattering and magnetization measurements are presented in the Supplemental Material~\cite{SM} (see, also, Ref. \onlinecite{Jiao_11,Jiao_13,Jeevan_11} therein for similar magnetization data in other Eu122 FMSCs). As presented below, superconducting and ferromagnetic transitions occur below $\mathit{T\rm_{sc}}$ = 22.5 K and $\mathit{T\rm_{C}}$ = 18.5 K, respectively. 

Figure 1 summarizes the SANS results obtained using the medium $Q$-range setting. As the background recorded at $\mathit{T}$ = 25 K in the paramagnetic state is subtracted, the displayed SANS scattering intensity around $q$ = 0 is of purely magnetic origin, arising from the scattering from large-scale ferromagnetic domains. Figs. 1(A) and 1(B) show the data at $\mathit{T}$ = 18 K and 5 K, respectively, measured in zero field ($\mathit{H}$ = 0 T) after a zero-field-cooling (ZFC) process. As shown in Fig. 1(C), a comparison between the rocking scans at these two temperatures integrated using the horizontal sector A or B clearly indicates an increasing intensity with decreasing temperature, due to a significant increase of the Eu$^{2+}$ moment at 5 K. In addition, the magnetic SANS intensity in zero field displays a star-shaped anisotropy with a four-fold symmetry, as shown in Figs. 1(A), 1(B), and S1B~\cite{SM}. Such an anisotropy is believed to be due to the pinning of ferromagnetic domains below $\mathit{T\rm_{C}}$ following the high-symmetry crystallographic directions, resulting in a wider coverage in real-space length scales along $a$/$b$ axes and a smaller coverage along the reciprocal $a^{*}$/$b^{*}$ directions~\cite{footnote}.  %By radially integrating the intensity within different sectors denoted with $1$ and $2$, respectively, as shown in Fig. 1(A), the $Q$-dependences of the magnetic intensity at 18 K for different azimuthal angles are plotted in Fig. 1(C) for the medium $Q$-range. Interestingly, the radially integrated intensity $I(Q)$ shows two characteristic humps on top of a linear slope corresponding to $I(Q)\propto Q^{-4}$, especially for that integrated using sector 1 orienting along the tetragonal [$\mathit{H}$, $\mathit{H}$, 0] and [$\mathit{H}$, $\mathit{-H}$, 0] directions.

Applying a longitudinal magnetic field parallel to the incident neutron beam along the $\mathit{c}$ axis effectively suppresses the magnetic SANS intensity, as shown in Figs. 1(D) and 1(E) at $\mathit{T}$ = 5 K, due to the suppression of in-plane ferromagnetic fluctuations around $q$ = 0 by the longitudinal field. In the meantime, the application of a longitudinal field rotates the ferromagnetic domains towards the field direction along $\mathit{c}$, leading to a homogenous azimuthal distribution of the in-plane length scale and a symmetric SANS pattern. In contrast, for a transverse field applied parallel to the $\mathit{ab}$ plane and perpendicular to the neutron beam, as shown in Fig. 1(F), a two-leaf anisotropy is observed as a result of the rotation of ferromagnetic domains towards the field direction along [1, -1, 0]. %, which is expected for diffuse magnetic SANS scattering from smooth ferromagnetic domains that are larger compared to the sensitivity of the SANS experiment ($\sim$ 300 nm). The complete sets of $q$-, temperature and field dependences of the magnetic SANS intensity will be further analyzed in detail below. 

\begin{figure}[b]
\includegraphics[width=0.5\textwidth]{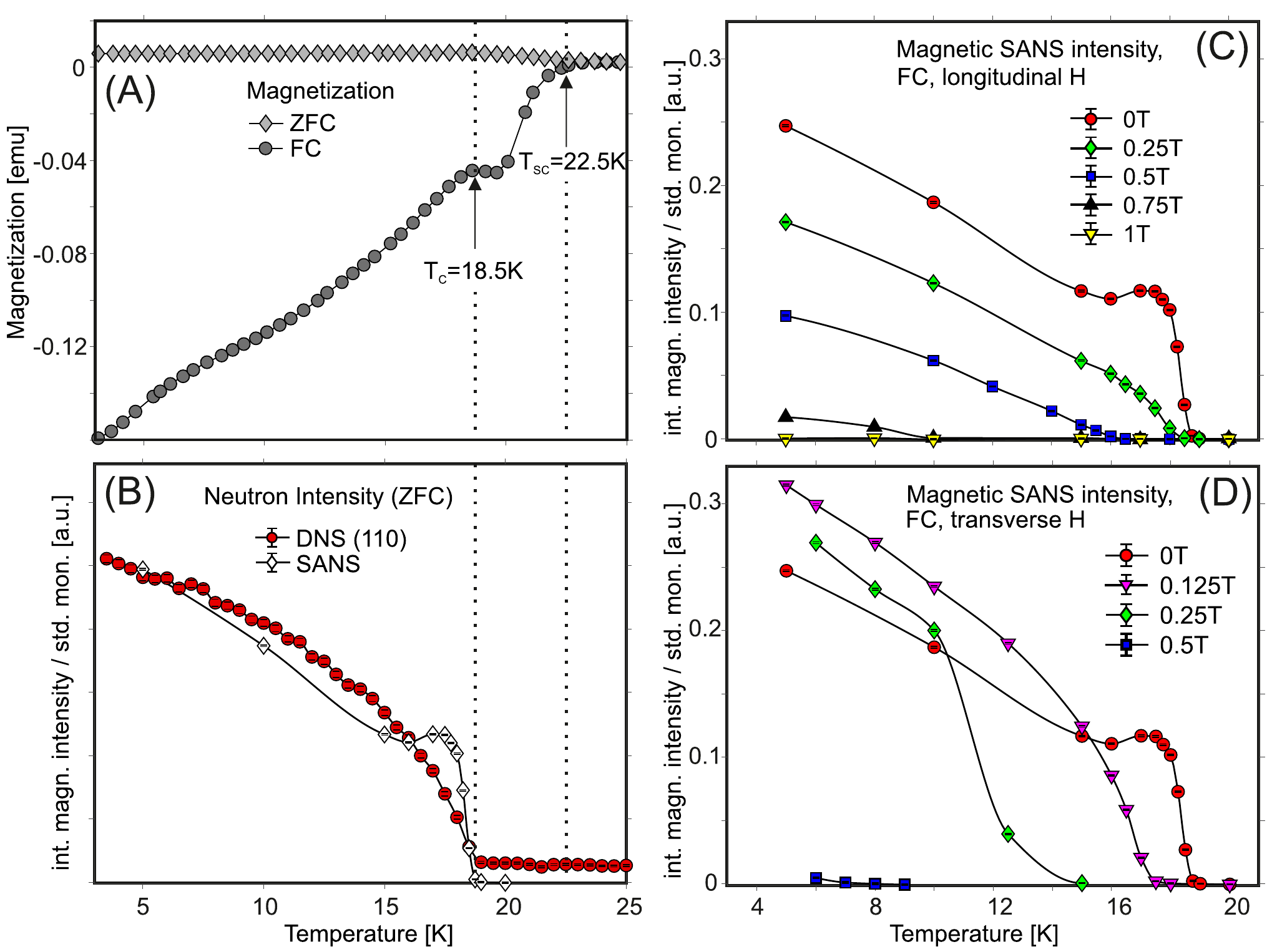}
\caption{DC magnetization of an EuFe$_{2}$(As$_{0.8}$P$_{0.2}$)$_{2}$ crystal measured in a magnetic field of 10 Oe parallel to the $\mathit{c}$ axis in ZFC and FC processes (A), the temperature dependences of the total integrated intensity of the (1 1 0) ferromagnetic reflection measured at the DNS spectrometer and the magnetic SANS signal (B) in zero field, as well as the effects of a longitudinal (C) and transverse (D) field on the magnetic SANS intensity. In (B), the two datasets are scaled to each other with respect to those at 5 K. In (C) and (D),  the temperature dependences are measured in FC processes under different fields.}
\label{Fig_2}
\vspace{-0.00\textwidth}
\end{figure}

Fig. 2(A) shows the DC magnetization data of one EuFe$_{2}$(As$_{0.8}$P$_{0.2}$)$_{2}$ crystal, which suggests the appearances of superconducting and ferromagnetic transitions below $\mathit{T\rm_{sc}}$ = 22.5 K and $\mathit{T\rm_{C}}$ = 18.5 K, respectively ~\cite{SM}. The temperature and field dependences of the integrated magnetic SANS intensity are presented in Figs. 2(B-D). The $\mathit{T\rm_{C}}$ value determined from the temperature dependence of magnetic SANS signal in zero field, as shown in Fig. 2(B), is perfectly consistent with those determined from the magnetization data and the zero-field neutron diffraction measurement, respectively. While the total integrated intensity of (1 1 0) Bragg reflection corresponding to the ferromagnetic ordering with a magnetic propagation vector of $k$ = 0 shows a typical Landau's order-parameter behavior, the magnetic SANS intensity is characterized by an additional hump around $\sim$ 17.5 K, associated with the appearance of the intermediate DMS, as will be discussed below, due to the interplay between FM and SC.  

Figs. 2(C) and 2(D) summarize the effects of longitudinal and transverse field, respectively, on the temperature dependences of integrated magnetic SANS intensity. The application of either a longitudinal or transverse field tends to suppress the intrinsic ferromagnetic fluctuations and accordingly the onset temperature of magnetic SANS scattering.  The SANS signal completely vanishes in a longitudinal field of $\mathit{H_{l}}$ = 1 T and a tranverse field of $\mathit{H_{t}}$ = 0.5 T. Furthermore, the hump exhibited close to $\sim$ 17.5 K in zero field almost disappears for $\mathit{H_{l}}$ = 0.5 T and $\mathit{H_{t}}$ = 0.125 T, respectively, suggesting the destructive role of the external field against the DMS.

\begin{figure*}
\begin{center}
\includegraphics[width=1\textwidth]{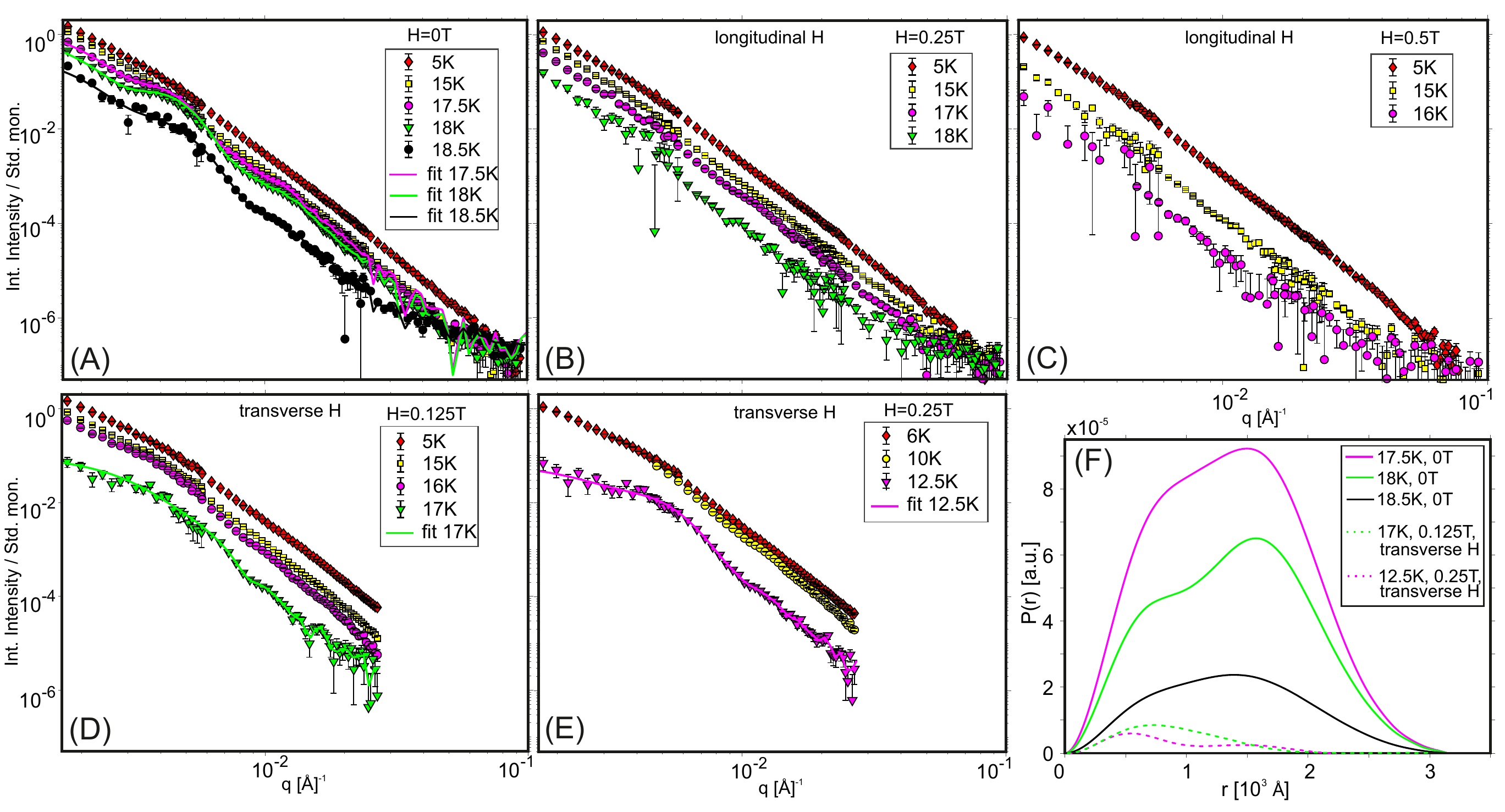}
\caption{$q$-dependences of the radially integrated magnetic SANS intensity at different temperatures in zero magnetic field (A), a longitudinal field of $\mathit{H_{l}}$ = 0.25 T (B) and $\mathit{H_{l}}$ = 0.5 T (C), a transverse field of $\mathit{H_{t}}$ = 0.125 T (D) and $\mathit{H_{t}}$ = 0.25 T (E), respectively. The $I(q)$ curves with a double-hump feature collected at the intermediate temperatures just below $T\rm_{C}$, for the zero-field and transverse-field case, were fitted using the IFF method, as the solid lines in (A), (D) and (E) represent. (F) shows the extracted real-space pair distance distribution function $P(r)$, revealing the existence of the DMS at intermediate temperatures closely below $T\rm_{C}$.}
\label{Fig_3}
\vspace{-0.00\textwidth}
\end{center}
\end{figure*}

The $q$-dependences of the magnetic SANS intensity of EuFe$_{2}$(As$_{0.8}$P$_{0.2}$)$_{2}$ at different temperatures, integrated using the white sectors as schematically sketched in Figs. 1(A), 1(D) and 1(F) for the zero-, longitudinal- and transverse-field cases, respectively, were summarized in Fig. 3. As shown in Fig. 3(A), in zero field, all the $I(q)$ curves for $\mathit{T}$ < $\mathit{T\rm_{C}}$ show an overall sloping background with the power-law behavior of $I(q)\propto Sq^{-4}$, known as the Porod's law expected for scattering from smooth ferromagnetic domains with flat interfaces independent of $q$ ($S$ is the domain wall area in our case)~\cite{Porod_51}. The increasing SANS intensity and vertical intercept of $I(q)$ curves with decreasing temperature indicates the increase of the ferromagnetic moment and the growth of ferromagnetic domains upon cooling. Interestingly, for the temperatures closely below $T\rm_{C}$ in an interval of $\sim$ 1.5 K, the $I(q)$ curves clearly exhibit a double-hump structure superimposed on a sloping background.
 
Using the indirect Fourier transform (IFF) method introduced in Ref. \onlinecite{Bender_17}, the real-space pair distance distribution function $P(r)$ in EuFe$_{2}$(As$_{0.8}$P$_{0.2}$)$_{2}$ for $T=17.5\,$K, $18\,$K and $18.5\,$K at $H=0$\,T can be extracted according to the relation
\begin{equation*}
I(q) = 4\pi\int_{0}^{\infty} P(r)\frac{\rm{sin}(qr)}{qr}\, \rm{d}r + \rm{background}
\end{equation*} and plotted in Fig. 3(F), where characteristic length scales in between $\sim$80 to $\sim$160 nm are identified to show up at these intermediate temperatures. These additional length scales revealed here are completely consistent with the previous observation of an intermediate inhomogeneous DMS using MFM, characterized by striped domains with the width in between 100 to 200 nm, in a very narrow temperature range ($\sim$ 1 K) just below $\mathit{T\rm_{C}}$ in EuFe$_{2}$(As$_{0.79}$P$_{0.21}$)$_{2}$~\cite{Stolyarov_18}. As shown by the solid curves in Fig. 3(A), the superimposition of these nanometer-scaled structures with larger smooth ferromagnetic domains fits the  $I(q)$ curves perfectly. Upon further cooling ($\mathit{T}$ $\ll$ $\mathit{T\rm_{C}}$), the double humps smear out when the FM is fully developed (see Fig. S1A for a comparison between 18 K and 5 K)\cite{SM}, indicating that these additional length scales exist only in an intermediate state residing in a very narrow temperature range below $T\rm_{C}$ and reflect the competition between FM and SC. Therefore, we regard them as convincing evidences of the appearance of DMS in EuFe$_{2}$(As$_{0.8}$P$_{0.2}$)$_{2}$, but more importantly, in a bulk form, as SANS is a bulk probe of large-scale structure and magnetism~\cite{Muehlbauer_19}. 

Futhermore, the $q$-dependences of the magnetic SANS intensity at different temperatures in longitudinal and transverse fields were plotted in Figs. 3(B-C) and 3(D-E), respectively.  In a longitudinal field of $\mathit{H_{l}}$ = 0.25 T (Fig. 3(B)) and $\mathit{H_{l}}$ = 0.5 T (Fig. 3(C)), the double-hump feature can still be indentified at intermediate temperatures, but less pronounced. However, the $I(q)$ behavior observed in a transverse field is qualitatively different, in which the double humps are basically replaced by a single hump. As shown in Fig. 3(F), fitting the $I(q)$ curves for $\mathit{H_{t}}$ = 0.125 T at $\mathit{T}$ = 17 K and $\mathit{H_{t}}$ = 0.25 T at $\mathit{T}$ = 12.5 K using the same IFF method yields a smaller length scale in between $\sim$ 50 to $\sim$ 70 nm, suggesting a more effective role of transverse field in killing the DMS into domains with even smaller sizes. The general tendency of $I(q)$ evolving with decreasing temperature in external fields is similar to the zero-field case, with the total magnetic SANS intensity increasing and the broad feature associated with nanometer-scaled DMS smearing out. 

\begin{figure}
\includegraphics[width=0.5\textwidth]{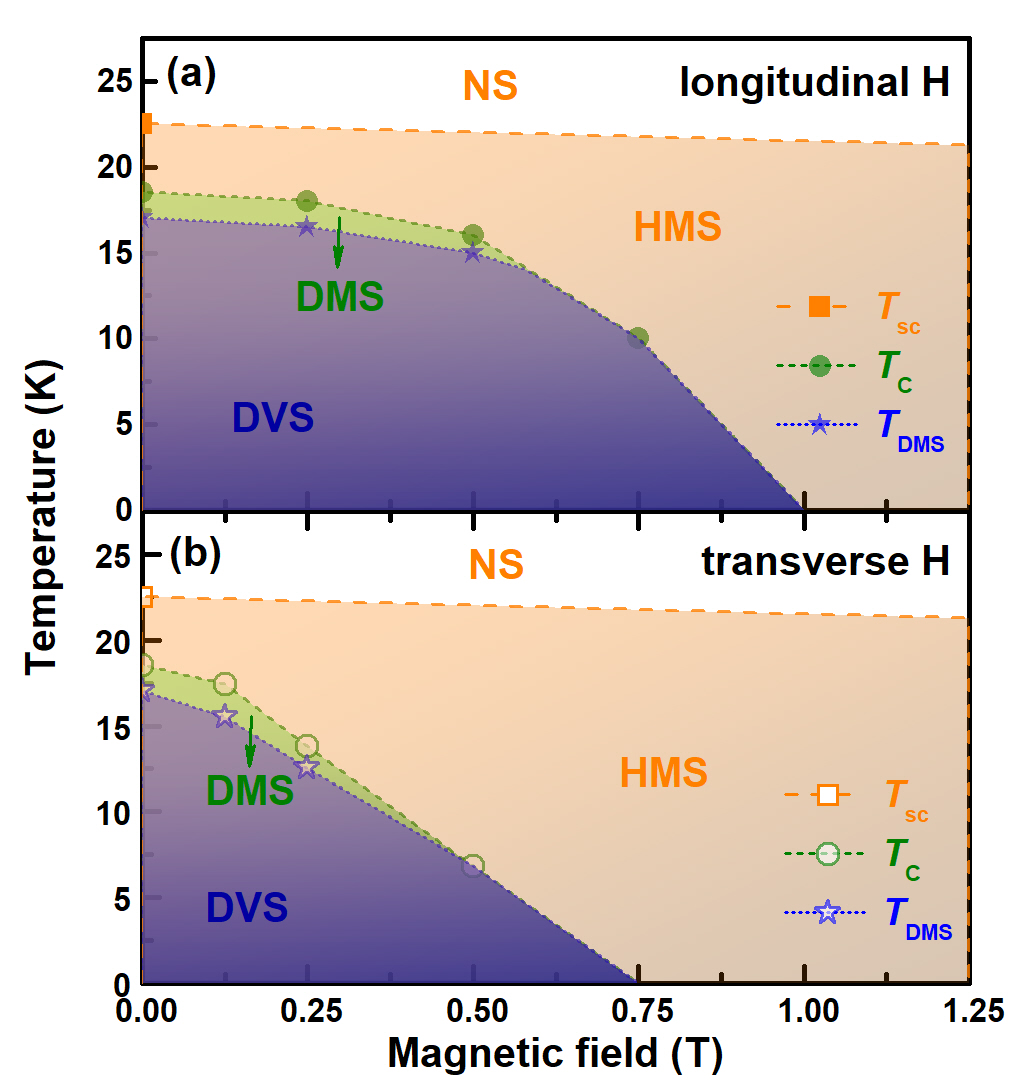}
\caption{Temperature-field phase diagrams for EuFe$_{2}$(As$_{0.8}$P$_{0.2}$)$_{2}$ under an applied longitudinal (a) and transverse (b) magnetic field, respectively. $T\rm_{C}$ denotes the temperature below which the ferromagnetic correlations get established. $T\rm_{sc}$ denotes the superconducting transition temperature, which is supposed to be suppressed very slowly under 2 T for EuFe$_{2}$(As$_{1-x}$P$_{x}$)$_{2}$ \cite{Nandi_14}. NS, HMS, DMS and DVS denote the normal state above $T\rm_{sc}$, the homogeneous Meissner state in between $T\rm_{sc}$ and $T\rm_{C}$, the domain Meissner state in between $T\rm_{C}$ and $T\rm_{DMS}$, and  the domain vortex-antivortex state below $T\rm_{DMS}$, respectively. $T\rm_{DMS}$ is determined according to the temperature at which the magnetic SANS intensity in Fig. 2(C,D) displays an additional feature and the $I(q)$ profiles shows a clearly identified double-bump structure.}
\label{Fig_4}
\vspace{-0.00\textwidth}
\end{figure}

In stark contrast to the previous SANS results on isostructural 122-family iron-based superconductors, such as KFe$_{2}$As$_{2}$ and doped BaFe$_{2}$As$_{2}$~\cite{Eskildsen_09,Demirdis_16,Kawano-Furukawa,Eskildsen_11}, where either a disordered or ordered vortex lattice was observed in the superconducting state, no evidence of Abrikosov vortices can be identified in EuFe$_{2}$(As$_{0.8}$P$_{0.2}$)$_{2}$ below $T\rm_{sc}$. Even for the temperatures in between $T\rm_{C}$ and $T\rm_{sc}$, where a homogeneous Meissner state is expected, the large background of strongly fluctuating FM of Eu together with the imperfection of coaligned crystals makes the detection of the superconducting vortex lattice almost impossible despite considerable counting efforts. However, a clear signature of the formation of smooth and large ferromagnetic domains outside the resolution range of the SANS instrument ($\sim$300 nm) is found below $T\rm_{C}=18.5\,$K, featured by characteristic scattering with $I(q)\propto q^{-4}$. This is well consistent with the observation of the domain vortex-antivortex state (DVS) characterized by ferromagnetic domains of large size ($\mathit{l}$ $\sim$ 350 nm) well below $T\rm_{C}$ by MFM technique \cite{Stolyarov_18}. 

Summarizing all results obtained in this study, the temperature-field phase diagrams for EuFe$_{2}$(As$_{0.8}$P$_{0.2}$)$_{2}$ under applied longitudinal and transverse fields are constructed and plotted in Fig. 4. Most importantly, in a small temperature interval of $\sim$ 1.5 K below $T\rm_{C}$ ($\mathit{T\rm_{DMS}}$ < $\mathit{T}$ < $\mathit{T\rm_{C}}$, where $\mathit{T\rm_{DMS}}$ denotes the temperature below which the double-hump feature associated with the DMS is no longer pronouncedly visible), when the ferromagnetic correlations are just established, additional SANS signals in form of double humps associated with previously proposed nanometer-scaled Meissner domains (DMS) is observed, suggesting that the decoration of SC on the FM in forms of additional large-scale domain structures is in fact of a bulk nature. When driven away from the narrow phase region of the DMS by lowering the temperature, these Meissner domains with the size of $\sim$ 100-200 nm are quickly buried into larger ferromagnetic domains when the FM gets enhanced. Further SANS measurements in magnetic fields in more details will be crucial for better understanding the effects of external fields on the evolution of length scales of the intermediate DMS and manipulating the interplay between the FM and SC.

In conclusion, by systematically investigating the magnetic SANS intensity of the FMSC EuFe$_{2}$(As$_{0.8}$P$_{0.2}$)$_{2}$ as a function of temperature and magnetic field, we confirm that the DMS observed in the previous surface-sensitive MFM measurement is factually of a bulk nature and exists only in a narrow temperature and magnetic field range as an intermediate state. Ascribing to the subtle interplay between FM and SC in the Eu122 FMSCs, these nanometer-scaled Meissner domains are finally buried into bulk ferromagnetic domains with much larger size upon cooling, corresponding to the DVS in the ground state. The temperature-field phase diagrams of EuFe$_{2}$(As$_{0.8}$P$_{0.2}$)$_{2}$ under applied longitudinal and transverse magnetic fields are constructed to describe the competition between SC and FM. Our results provide a key solution to the mystery regarding the intriguing coexistence of strong ferromagnetism and bulk superconductivity in the Eu122 FMSCs. 

\bibliographystyle{apsrev} \bibliographystyle{apsrev}

\begin{thebibliography}{10}
\bibitem{footnote} We note that the background at $\mathit{T}$ = 25 K also shows such an anisotropy with four-fold symmetry, most likely due to the intrinsic crystal defects or reflections by the stacking faults.
\end{thebibliography}
\begin{acknowledgments}
We would like to acknowledge Susanne Mayr for the assistance with the co-alignment of the crystals, and Wenhe Jiao for stimulating discussions. This work is based on experiments performed at the SANS-1 instrument operated by Technische Universit\"at M\"unchen and DNS instrument operated by J\"ulich Centre for Neutron Science (JCNS) at the Heinz Maier-Leibnitz Zentrum (MLZ), Garching, Germany. This work was supported by the National Natural Science Foundation of China (Grant No. 12074023) and National Key Research and Development Program of China (2016YFA0300202), and the Fundamental Research Funds for the Central Universities in China.

\end{acknowledgments}

\end{document}